# Protein function influences frequency of encoded regions containing VNTRs and number of unique interactions.

Suzanne Bowen[1,*]

[1] Birkbeck College, University of London, Malet Street, Bloomsbury, London, WC1E 7HX

*Corresponding author, e-mail: sbowen02@mail.bbk.ac.uk

Running title: Influence of genomic stability on protein interactions.

**Abstract**

Proteins encoded by genes containing regions of variable number tandem repeats (VNTRs) are known to be polymorphic within species but the influence of their instability in molecular interactions remains unclear. VNTRs are overrepresented in encoding sequence of particular functional groups where their presence could influence protein interactions. Using human consensus coding sequence, this work examines if genomic instability, determined by regions of VNTRs, influences the number of protein interactions. Findings reveal that, in relation to protein function, the frequency of unique interactions in human proteins increase with the number of repeated regions. This supports experimental evidence that repeat expansion may lead to an increase in molecular interactions. Genetic diversity, estimated by $K_a/K_s$, appeared to decrease as the number of protein-protein interactions increased. Additionally, G+C and CpG content were negatively correlated with increasing occurrence of VNTRs. This may indicate that nucleotide composition along with selective processes can increase genomic stability and thereby restrict the expansion of repeated regions. Proteins involved in acetylation are associated with a high number of repeated regions and interactions but a low G+C and CpG content. While in contrast, less interactive membrane proteins contain a lower number of repeated regions but higher levels of C+G and CpGs. This work provides further evidence that VNTRs may provide the genetic variability to generate unique interactions between proteins.

**INTRODUCTION**

Tandem repeats (TRs) in encoding DNA are unstable and prone to polymorphisms (Verkerk *et al.*, 1991; Bowen *et al.*, 2005; Usdin, 2008). Mutation rates of TRs within the genomes of various organisms can be over 10,000 times higher than in other parts of the genome; in humans the rate is around $10^{-4}$ mutations per locus per generation (Brinkmann *et al.*, 1998; Ellegren, 2004; Legendre *et al.*, 2007). As these genes are susceptible to genetic variability and vulnerable to deleterious mutations it might be assumed that their products would not be good candidates for reliable protein-protein interactions (PPIs). However, VNTRs seem to be preserved through selection despite their genetic instability (Mularoni *et al.*, 2012). Nearly half of the human genome comprises repeated sequences of either transposable elements, which include interspersed nuclear elements (LINEs and SINEs), or VNTRs. VNTRs can be further characterised into those that encode homopeptide sequence that arise from trinucleotide repeats (Faux *et al.*, 2005) or those that encode oligopeptide repeats, usually referred to as minisatellites (Jeffreys *et al.*, 1985; Richard and Dujon, 2006).

TRs were once viewed as being generated erroneously with no specific purpose (Smith, 1976). Now it has been proposed that they may bestow an adaptive advantage to cells (Gemayel *et al.*, 2010; Mularoni *et al.*, 2012). Intrinsically unstructured regions of proteins, such as those found in prions and RNA polymerase, are thought to evolve through repeat expansion (Tompa, 2003). Many repeats have been found to encode domains with a diverse range of functions such as metal binding and glycosylation. Altering the number of TRs in yeast flocculin proteins can influence cell adhesion through increasing glycosylation sites (Verstrepen *et al.*, 2005). Dynamic mutations, through TR variation, may contribute to adaptation or rapid evolution in lower organisms but in vertebrates they have also been implicated in genetic disease (Usdin, 2008). A number of neurological conditions are attributed to TR instability, including Huntington disease (HD) (Andrew *et al.*, 1993), where repeats are located within exons, and Fragile X Syndrome (Verkerk *et al.*, 1991), where they exist outside coding regions. It has been found that expansion of the polyglutamine repeats within the HD gene product can result in increased interactions with proteins related to specific functions such as energy production, protein trafficking, RNA post-transcriptional modifications and cell death (Ratovitski *et al.*, 2012). Although the consequence of TR instability in neurological genes may be detrimental in some cases there is also the possibility that polymorphisms could favourably increase novel interactions to offer an adaptive advantage. For instance, in canine genomes, repeats are found in developmental genes which have been associated with morphological plasticity (Fondon and Garner, 2004).

Repeat variability at the genomic level is thought to arise through various DNA repair processes involving nonhomologous recombination or replication slippage (Calabrese *et al.*, 2001; Liu and Wilson, 2012). Although a degree of instability may be advantageous, some kind of regulatory process may exist to prevent DNA from becoming too unstable. One such stabilising process could be methylation. Processes involving methylation, whereby a methyl group is added to CpG dinucleotides, are thought to lead to the suppression of repetitive elements and it has been found that demethylation increases the frequency of genome-wide trinucleotide repeat expansion (Dion *et al.*, 2008). Trinucleotide repeats are thought to owe their instability via secondary structures in single-stranded DNA so perhaps methylation could in some instances prevent these structures from occurring (Nichol and Pearson, 2002). The expansion of repeats could result in elevated binding of a specific protein or molecule but how likely are they to encourage the generation of novel interactions. This work examines if a relationship exists between TR expansion, genomic stability and novel interactions in proteins.

**RESULTS**

An over-representation of proteins encoded by genes containing VNTRs exist in various functional categories including transcription ($p=3.68 \times 10^{-09}$), DNA binding ($p=1.15 \times 10^{-11}$), epidermal development ($p=2.33 \times 10^{-11}$), collagen ($p=6.61 \times 10^{-09}$) and metal binding ($p=2.91 \times 10^{-05}$). This study used functional groups with significant counts ($p<0.05$) to evaluate if protein interactions are influenced by the presence of VNTRs.

**Protein interactions associated with frequency of repeated regions.**

Overall, 27,405 unique interactions were found in the BioGRID database for proteins encoded by DNA containing TRs. A network visualisation was produced from gene lists in functional categories with significant interactions (p-value<0.05) (Fig. 1). Pearson moment product correlation indicated that an association exists between repeated regions in CCDS and the number of PPIs ($r=0.0619$; $p=0.0043$; d.f. 2,116)). Placing genes into functional groups allows this correlation to be viewed graphically ($r=0.8321$; $p=0.0008$; d.f. 10)(Fig. 2). This association possibly reflects the importance of repeated regions to specific functional properties. Significant differences occur in the number of interactions associated with specific functional groups. For instance, a higher number of interactions occur in proteins involved in acetylation ($n=262$) compared to those that were located in the membrane ($n=295$) ($\bar{x}=22.68$, 4.67, $p=1.4 \times 10^{-13}$). There was also a higher number of TRs in acetylation proteins compared to membrane proteins ($\bar{x}=1.82$, 2.67, $p=1.10 \times 10^{-4}$) but no significant difference in gene length between these functional groups ($\bar{x}=2.89$kb, 2.91kb, $p=0.9224$)(Fig. 2).

No association was found between gene length and the number of regions containing TRs in genes within functional groups ($r=-0.2366$; $p=0.4837$; d.f. 10). Neither was there any association between the length of TRs and the number of repeated regions ($r=0.3888$; $p=0.2117$; d.f. 10). Curiously, there was a negative correlation between the length of repeats and that of genes ($r$= -0.6783; $p$= 0.0153; d.f. 10)(Fig. 2) but no association exists between repeat length and protein interactions ($r$= 0.3259; $p$= 0.3011; d.f. 10). Therefore, it is the number of repeated regions rather than their combined length that results in an elevated association with PPIs. Furthermore, the combined length of repeats does not necessarily contribute to gene length. The length of repeated regions seems to be influenced by protein function (Fig. 2). In metal binding proteins (n=615) repeat length is longer ($\bar{x}$=360.17, 118.23, $p$=<1.00x10$^{-15}$) than in RNA binding proteins (n=308) but in contrast mean gene length is shorter ($\bar{x}$=2490.21, 3725.51, $p$=<1.00x10$^{-15}$), perhaps reflecting a different mechanism of binding between these two functional groups.

**Influence of genetic stability on frequency of protein interactions.**

Having established that repeat expansion could lead to a greater number of novel interactions, the contribution of selection pressure in this process was investigated. Genetic stability as a consequence of SNP variation in genome 1000 data were used to determine the non synonymous/synonymous ($K_a/K_s$) ratio in proteins encoded by genes containing TRs. Pearson moment correlation revealed that positive selection decreased as protein interactions increase when proteins are classified in functional groups ($r$ = -0.9465; p=3.15x10$^{-6}$; d.f. 10) (Fig. 3). This indicates that selection is stronger in proteins that have a higher number of unique interactions and that function is probably an important factor in this process. When function is excluded from the analysis, the same correlation does not exist ($r$= 0.0241; $p$=0.2681; d.f. 2,116). Particular functional groups interact with more proteins and these proteins are subjected to greater selection pressure. For instance, as previously stated, a significantly higher number of unique interactions occur in proteins involved in acetylation compared to membrane located proteins. There also exists a significant difference between $K_a/K_s$ values between these two protein groups, with acetylation being lower ($\bar{x}$=1.7735, 2.9431, $p$=0.0268). Therefore, as repeat expansion is related to protein function so is the frequency of protein interactions and rate of selection. These results indicate that there is a relationship between protein interactions with regards to function and evolution within genes containing repeats.

Repeat stability has been associated with a number of factors including the entropy of the repeat sequence, the number of repeats within a tract and the direction of replication (Richard *et al*., 2008). Investigators have also discovered that methylation may be a contributing factor to repeat polymorphisms (Nichol and Pearson, 2002). As CpG dinucleotides have been associated with increased repeat stability this study investigated if there was any connection with these and the number of repeated regions or if either G+C or CpG content influenced the frequency of PPIs. To establish if repeat nucleotide composition could influence unique PPIs, G+C and CpG content were determined within genes and TRs. G+C content was significantly higher in TRs than in regions outside repeats ($\bar{x}$=61.92, 49.60, $p$=<1x10$^{-15}$). Unsurprisingly, so was CpG content ($\bar{x}$=5.20, 3.54, $p$=<1x10$^{-15}$). Similar findings have been observed in other studies (Saxonov *et al*., 2006). Interestingly, a large proportion of the genes (26.73%) contained no CpGs within TRs. Even though 26.73% of genes contained repeated regions with no CpG content, repeat CpG content was significantly higher than CpG content outside repeated regions ($\bar{x}$=3.54%, 5.20%, $p$=<1.00x10$^{-15}$).

There is a negative correlation between G+C% of genes and PPIs in functional groups ($r$=-0.7948; $p$=0.0021; d.f. 10). Similarly, there is an association between PPI and CpG content ($r$=-0.6935; $p$=0.0124; d.f. 10). There was a significantly lower number of CpGs in acetylation compared to membrane proteins, these functional groups had maximum and minimum CpG content, ($\bar{x}$= 3.61%, 4.63%, $p$=5.7x10$^{-7}$). TRs were analysed separately for CpG content (Fig. 4). Genes and TRs differed in distribution with CpG content in repeats having far higher variability than regions outside repeats with a standard deviation of 6.77 compared to 2.94 and sample variance of 45.87 compared to 5.75 ($n$=2,118) (Fig. 4).

Generally, there was no association between the number of repeats and G+C content ($r$=0.0241; $p$=0.2677; d.f. 2,116). There was a slightly better association between G+C content and the number of repeats when placed into functional groups but this was still not significant ($r$=-0.5121; $p$=0.0887; d.f. 10). A stronger correlation was observed between CDSs containing repeats and CpG content ($r$=-0.7490; $p$=5.77x10$^{-4}$; d.f. 10). This study confirms that as CpG content decreases the number of repeats increase. This gives the impression that CpG content rather than overall G+C content may in some way increase the stability of VNTR regions, perhaps in preventing homologous recombination which in turn suppresses the expansion of repeats.

**DISCUSSION**

Repeat expansion is thought to contribute to protein evolution by modifying interaction sites or enabling novel functions (Tompa, 2003). This study provides further evidence that repeated regions my supply the variability for proteins to adapt or increase their associative networks. Neurological proteins prone to repeat expansion attract a great deal of research because they are known to give rise to disease, such as Huntington's (HD) (Andrew *et al*., 1993; Verkerk *et al*., 1991). In fact, expansion of the polyglutamine repeats within the HD gene product has been shown to increase protein interactions (Ratovitski *et al*., 2012). Perhaps the predisposition for these regions to be unstable may also be beneficial as far as adaptation is concerned. Other studies have proposed that TRs within the human genome could promote genome reorganization during evolutionary processes (Ruiz-Herrera *et al*., 2006; Farré *et al*., 2011). Ruiz-Herrera *et al* (2006) discovered that human genomes contain a greater number of TRs in evolutionary breakpoint regions than other great ape species. VNTRs seem to provide the dynamic diversity on which selection can act and therefore may give humans an adaptive advantage.

Nucleotide composition also seems to influence the stability of TRs as the number of CpG regions in genes decrease with the number of protein interactions and repeated regions. Methylation may prevent repeat expansion which in turn would limit the number of protein interactions. There is conflicting information about whether methylation does in fact influence repeat stability in humans. In fungal species methylation has been shown to influence homologous recombination (HR) but it has been difficult to replicate results in mammalian cells (Dion *et al*., 2008). In CHO cells, demethylation, induced by treatment with 5-aza-2′-deoxycytidine, can lead to the expansion of trinucleotide repeats but does not seem to alter the frequency of HR. GC rich TRs have also been implicated in the binding of the ATRX gene product which has been associated with TR-related diseases such as X-linked mental retardation (XLMR) syndrome (Law *et al*., 2010). It is not clear what the function of the ATRX gene product is, but it has been implicated in gene regulation and methylation (Gibbons and Higgs, 2010). It has also been proposed that it may be involved in the recognition of unusual forms of DNA and assist in their resolution (Law *et al*., 2010). Indeed. ATRX associates with NEK1 which in turn interacts with MRE11, an endonuclease that forms part of the MRN/X complex, which associates with TRs in yeast genomes (Bowen and Wheals, 2006). The presence of methylated DNA may prevent homologous recombination or the formation of secondary structures and thereby limit the expansion of TRs.

Alternatively, GC rich regions in TRs could just reflect selective processes acting on the physical properties of the protein and may have little connection with genome stability. It is well documented that amino acid repeats share common properties that are associated with function (Albà *et al*., 1999; Mularoni *et al*., 2010). For instance, physical characteristics of amino acids in membrane proteins contribute to their overall function. Amino acids that interact with the lipid bilayer tend to be hydrophobic in order to repel reactive carbonyl and amine peptide bonds, whereas those that protrude from the surface of the membrane are usually hydrophilic. Compared to membrane proteins, acetylation-related proteins are far more interactive, as eighty percent of human proteins are acetylated, including transcription factors, effector proteins, molecular chaperones, and cytoskeletal proteins. Acetylation is a post-translational modification analogous to phosphorylation. There are a significantly greater number of repeats and PPIs in proteins involved in acetylation compared to those in the membrane. Recently it has been proposed that acetylated proteins could influence other regulatory processes, such as, phosphorylation, methylation, ubiquitination and sumoylation (Yang and Seto, 2008). The genes belonging to this functional group also had the lowest G+C and CpG content.

Overall this study illustrates that the function of proteins influences the frequency of VNTR regions and that this in turn may lead to an increased number of unique PPIs that could be influenced by selection processes and nucleotide composition.

**METHODS**

Encoding nucleotide sequence (approx 26,500 genes) from the Ensembl consensus coding DNA sequence project (CCDS) in FASTA format were downloaded from the NCBI ftp site (Pruitt *et al*., 2009, ftp://ftp.ncbi.nlm.nih.gov/pub/CCDS/) then screened for the presence of repeats using Tandem Repeat Finder with match, mismatch, and indel weightings of 5, 5, and 7, minimum alignment score of 50 and maximum period size of 500 (Benson *et al*., 1999). Output of FASTA files that masked the presence of detected repeats were also created. Using masked and unmasked files a regular expression Perl script computed the G+C content in genes and repeats thereby determining contribution of repeats to overall G+C composition. A sliding window Perl script determined regions that contained CG dinucleotides.

Genes containing regions of TRs (*n*=4,268) were placed into functional groups using data from the Gene Ontology Database (The Gene Ontology Consortium, 2000; http://geneontology.org). CCDSs were allocated into functional clusters using a fuzzy heuristic partition algorithm (Huang *et al*., 1999, http://david.abcc.ncifcrf.gov/home.jsp). As proteins can contribute to more than one function, this algorithm allows genes to be placed in relevant classifications, with outliers being excluded from the data. A Fisher Exact test was used to measure gene-enrichment in annotation terms. Classifications containing a significant proportion of genes were further analysed for unique interactions.

PPIs were determined for each gene from data downloaded from BioGRID (Stark *et al*., 2011; http://thebiogrid.org) and uploaded into a mySQL database to determine relational associations. The mySQL database contained entries for 2,118 unique proteins encoded by genes containing repeated regions. These were involved in 27,405 interactions within twelve functional group nodes. Network visualisations between functional groups were created by applying a Kamada-Kawai algorithm to gene clusters to create a network of nodes involved in significant ($p<0.001$) interactions (Shannon *et al*., 2003; Kaimal *et al.*, 2010). Genes encoding proteins involved in interactions were analysed further for genetic variation using SNP and INDEL calls from phase 1 integrated release of 1092 individuals and Ensembl release 65 from the 1000 Genomes Project (The 1000 Genomes Project Consortium, 2010; ftp://ftp.1000genomes.ebi.ac.uk/vol1/ftp).

**REFERENCES**

Albà MM, Santibáñez-Koref MF, Hancock JM (1999) Amino acid reiterations in yeast are over-represented in particular classes of proteins and show evidence of a slippage-like mutational process. *J. Mol. Evol.,* **49**: 789–797.

Andrew SE, Goldberg YP, Kremer B, Telenius H, Theilmann J *et al*., (1993) The relationship between trinucleotide (CAG) repeat length and clinical features of Huntington's disease. *Nat. Genet.*, **4**: 398–403.

Benson G (1999) Tandem repeats finder: a program to analyze DNA sequences. *Nucleic Acids Res.*, **27**: 573–580.

Bowen S, Roberts C, Wheals AE (2005) Patterns of polymorphism and divergence in stress-related yeast proteins. *Yeast*, **22**: 659–668.

Bowen S and Wheals AE (2006) Evidence that protein expansion and contraction is partly owing to pre-meiotic mutational events. *Mol. Biol. Evol*., **23**: 1339-1340.

Brinkmann B, Klintschar M, Neuhuber F, Huhne J, Rolf B (1998) Mutation rate in human microsatellites: influence of the structure and length of the tandem repeat. *Am. J. Hum. Genet.*, **62**: 1408–1415.

Calabrese PP, Durrett RT, Aquadro C F (2001) Dynamics of microsatellite divergence under stepwise mutation and proportional slippage/point mutation models. *Genetics*, **159**: 839–852.

Dion V, Lin Y, Price BA, Fyffe SL, Seluanov A *et al*. (2008) Genome-wide demethylation promotes triplet repeat instability independently of homologous recombination. *DNA Repair (Amst),* **7**: 313–320.

Ellegren H (2004) Microsatellites: simple sequences with complex evolution. *Nat. Rev. Genet.*, **5**: 435–445.

Farré M, Bosch M, López-Giráldez F, Ponsà M, Ruiz-Herrera A (2011) Assessing the role of tandem repeats in shaping the genomic architecture of great apes. PLoS One, 6:e27239.

Faux NG, Bottomley SP, Lesk AM, Irving JA, Morrison JR *et al*. (2005) Functional insights from the distribution and role of homopeptide repeat-containing proteins. *Genome Res.*, **15**(4): 537-551.

Fondon JW, Garner HR (2004) Molecular origins of rapid and continuous morphological evolution. *Proc. Natl. Acad. Sci*., **101**: 18058–18063.

Gemayel R, Vinces MD, Legendre M, Verstrepen KJ (2010) Variable tandem repeats accelerate evolution of coding and regulatory sequences. *Annu. Rev. Genet*., **44**: 445–477.

Gibbons RJ and Higgs DR (2010) ATRX: taming tandem repeats. *Cell Cycle*, **9**(23): 4605-4606.

Huang DW, Sherman BT, Lempicki RA (2009) Systematic and integrative analysis of large gene lists using DAVID Bioinformatics Resources. *Nature Protoc*., **4**: 44-57.

Jeffreys AJ, Wilson V, Thein SL (1985) Hypervariable 'minisatellite' regions in human DNA. *Nature*, **314**: 67-73.

Kaimal V, Bardes EE, Tabar SC, Jegga AG, Aronow BJ (2010) ToppCluster: a multiple gene list feature analyzer for comparative enrichment clustering and network-based dissection of biological systems. *Nucleic Acids Res*., **38**: W96-102.

Law MJ, Lower KM, Voon HP, Hughes JR, Garrick D *et al*., (2010) ATR-X syndrome protein targets tandem repeats and influences allele-specific expression in a size-dependent manner. *Cell*, **143**: 367-378.

Legendre M, Pochet N, Pak T, Verstrepen KJ (2007) Sequence-based estimation of minisatellite and microsatellite repeat variability. *Genome Res.*, **17**: 1787–1796.

Liu Y, Wilson SH (2012) DNA base excision repair: a mechanism of trinucleotide repeat expansion. *Trends Biochem. Sci.*, **37**: 162-172.

Mularoni L, Ledda A, Toll-Riera M, Albà MM (2010) Natural selection drives the accumulation of amino acid tandem repeats in human proteins. *Genome Res.*, **20**(6): 745-754.

Nichol K, Pearson CE (2002) CpG methylation modifies the genetic stability of cloned repeat sequences. *Genome Res.,* **12**: 1246-1256.

Pruitt KD, Harrow J, Harte RA, Wallin C, Diekhans M *et al*., (2009) The consensus coding sequence (CCDS) project: Identifying a common protein-coding gene set for the human and mouse genomes. *Genome Res.*, **19**: 1316-1323.

Ratovitski T, Chighladze E, Arbez N, Boronina T, Herbrich S, Cole RN, Ross CA (2012) Huntingtin protein interactions altered by polyglutamine expansion as determined by quantitative proteomic analysis. *Cell Cycle*, **11**: 2006-2021.

Richard GF, Dujon B (2006) Molecular evolution of minisatellites in hemiascomycetous yeasts. *Mol. Biol. Evol.*, **23**: 189–202.

Richard GF, Kerrest A, Dujon B (2008) Comparative genomics and molecular dynamics of DNA repeats in eukaryotes. *Microbiol. Mol. Biol. Rev.*, **72**: 686–727.

Ruiz-Herrera A, Castresana J, Robinson TJ (2006) Is mammalian chromosomal evolution driven by regions of genome fragility? *Genome Biol.*, **7**: R115.

Saxonov S, Berg P, Brutlag DL (2006) A genome-wide analysis of CpG dinucleotides in the human genome distinguishes two distinct classes of promoters. *P. Natl. Acad. Sci. USA.*, **103:** 1412–1417.

Shannon P, Markiel A, Ozier O, Baliga NS, Wang JT, Ramage D, Amin N, Schwikowski B, Ideker T (2003) Cytoscape: a software environment for integrated models of biomolecular interaction networks. *Genome Res.*, **13**: 2498-2504.

Smith GP (1976) Evolution of repeated DNA sequences by unequal crossover. *Science,* **191**: 528-535.

Stark C, Breitkreutz BJ, Chatr-Aryamontri A, Boucher L, Oughtred R *et al*., (2011) The BioGRID Interaction Database: 2011 update. *Nucleic Acids Res.*, D698-704.

Tompa P (2003) Intrinsically unstructured proteins evolve by repeat expansion. *BioEssays*, **25**: 847–855.

The Gene Ontology Consortium (2000) Gene ontology: tool for the unification of biology. *Nat. Genet.*, **25**: 25-29.

The 1000 Genomes Project Consortium (2010) A map of human genome variation from population-scale sequencing. *Nature*, **467**: 1061-1073.

Usdin K (2008) The biological effects of simple tandem repeats: Lessons from the repeat expansion diseases. *Genome Res.*, **18**: 1011-1019.

Verkerk AJ, Pieretti M, Sutcliffe JS, Fu YH, Kuhl DP *et al*., (1991) Identification of a gene (FMR-1) containing a CGG repeat coincident with a breakpoint cluster region exhibiting length variation in fragile-X syndrome. *Cell*, **65**: 905–914.

Verstrepen KJ, Jansen A, Lewitter F, Fink GR (2005) Intragenic tandem repeats generate functional variability. *Nat. Genetics*, **37**: 986–990.

Yang XJ, Seto E (2008) Lysine acetylation: codified crosstalk with other posttranslational modifications. *Mol. Cell*, **31**(4): 449–461.

**ACKNOWLEDGMENTS**

Many thanks to Alan Wheals for editing the manuscript.

**FIGURE LEGENDS**

Figure 1. Network featuring functional groups of proteins encoded by genes containing VNTRs involved in PPIs. Visualisations between functional groups (colour coded) were created by applying a Kamada-Kawai algorithm to gene clusters to create a network of nodes (Shannon *et al.*, 2003; Kaimal *et al.*, 2010). Only nodes involved in significant (p<0.001) interactions are shown. Functional group clusters were created from Gene Ontology data (The Gene Ontology Consortium, 2000; http://geneontology.org).

Figure 2. Association between VNTR frequency and PPIs. Overall length (in nucleotides) of genes and repeats (A), have less association with the number of PPIs than the frequency of repeated regions (B). Mean ± SEM.

Figure 3. Relationship between frequency of PPIs and genetic stability. Interactions increase with the frequency of repeated regions (A) but decrease in response to selection pressure (B). Frequency of repeated regions decrease as CpG content increases (C) and genetic diversity decreases as the frequency of repeated regions increases (D). This implies that selection pressure limits repeat expansion and the number of PPIs.

Figure 4. Distribution of CG dinucleotides within genes and repeated regions. A large number of repeated regions (26.73%) do not contain CpGs but overall frequency is higher within repeats than in regions outside repeats ($\bar{x}$=3.54%, 5.20%, *p*=<1.00x10$^{-15}$).

Figure 1

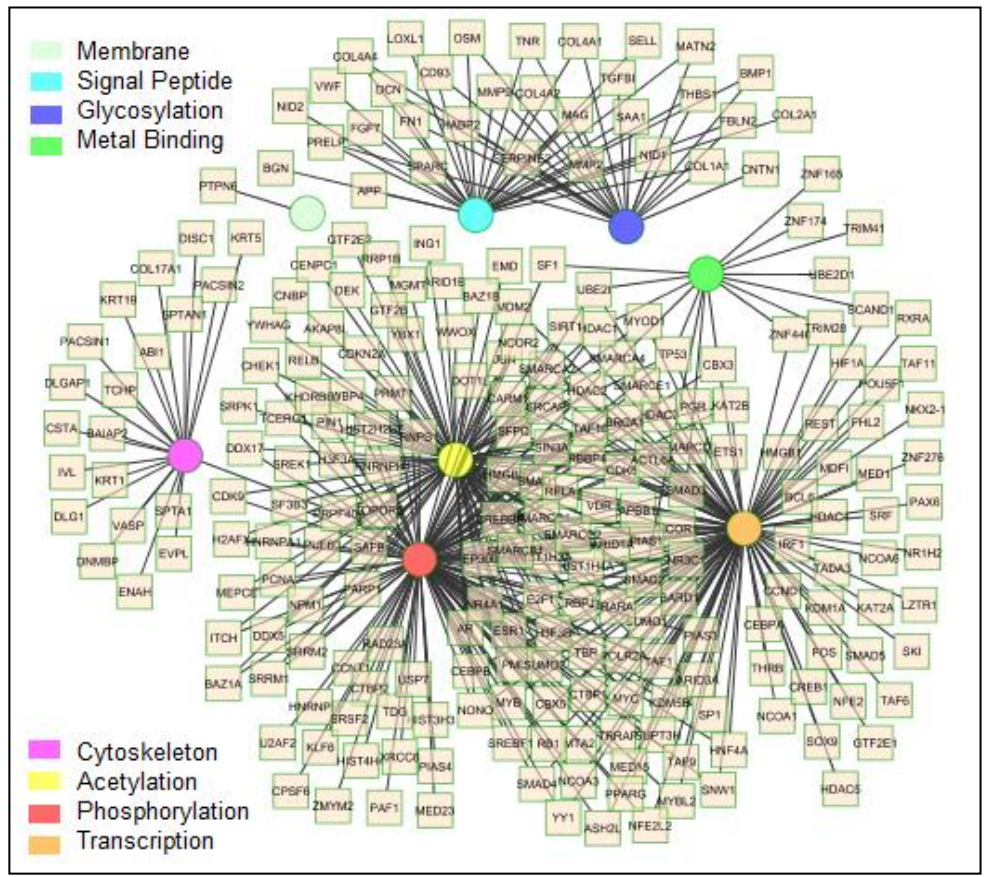

Figure 2

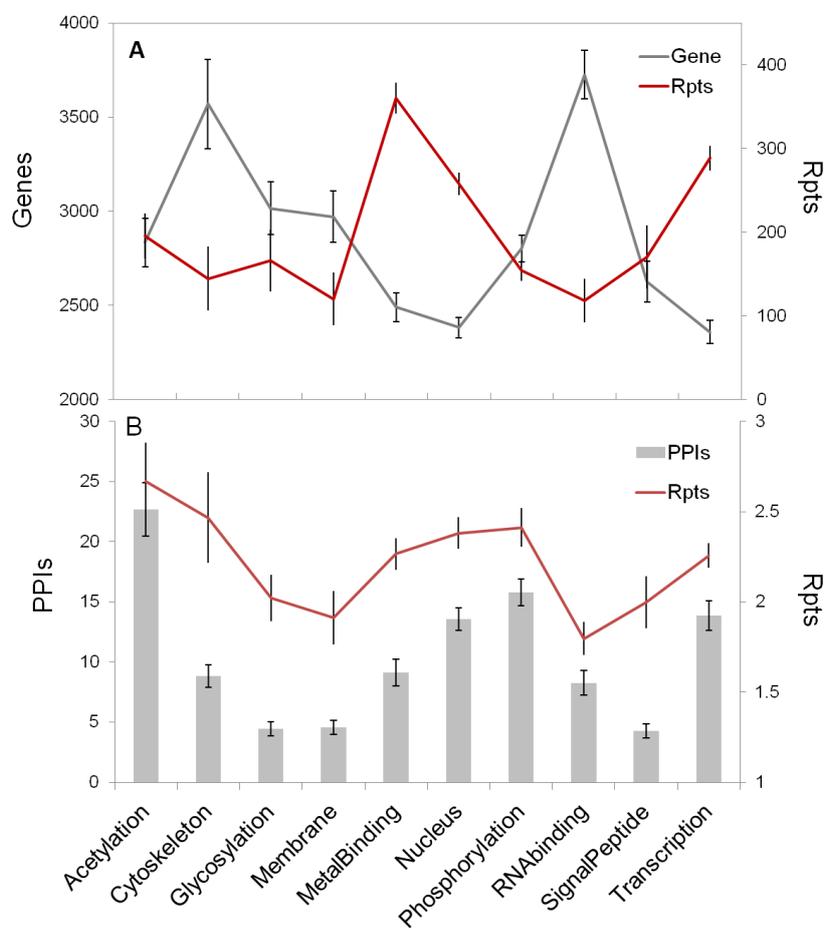

Figure 3

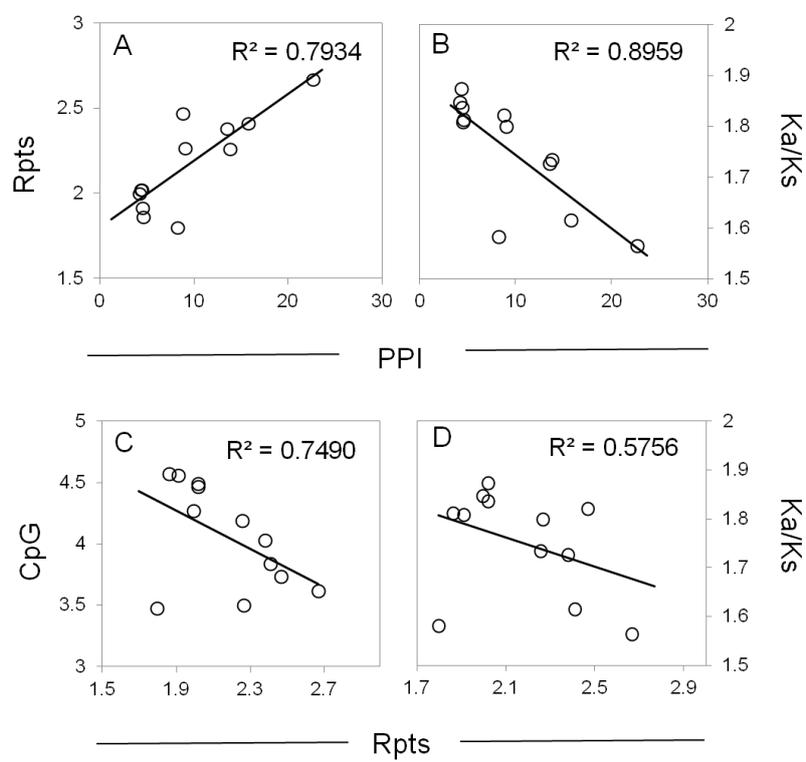

Figure 4

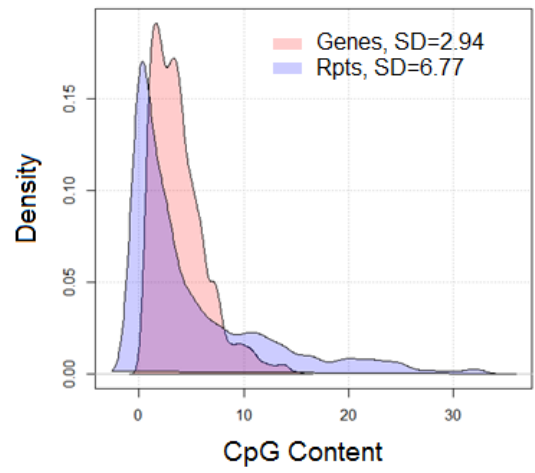